**Room-temperature observation and current control of skyrmions in Pt/Co/Os/Pt thin films**

R. Tolley[1,2], S. A. Montoya[3], and E. E. Fullerton[1,2] *

[1]*Center for Memory and Recording Research, University of California, San Diego, La Jolla, CA 92093-0401, USA*
[2]*Department of Electrical and Computer Engineering, University of California, San Diego, La Jolla, CA 92093, USA*
[3]*Space and Naval Warfare Systems Center Pacific, San Diego, CA 92152, USA*

Dated: November 3, 2017

**Abstract.**

We report the observation of room-temperature magnetic skyrmions in Pt/Co/Os/Pt thin-film heterostructures and their response to electric currents. The magnetic properties are extremely sensitive to inserting thin Os layers between the Co-Pt interface resulting in reduced saturation magnetization, magnetic anisotropy and Curie temperature. The observed skyrmions exist in a narrow temperature, applied-field and layer-thickness range near the spin-reorientation transition from perpendicular to in-plane magnetic anisotropy. The skyrmions have an average diameter of 2.3μm and transport measurements demonstrate these features can be displaced with current densities as low as $J = 2 \times 10^4$ A/cm$^2$ and display a skyrmion Hall effect.

* Corresponding author: efullerton@ucsd.edu

Skyrmions are topologically protected magnetic domain structures that have attracted considerable interest given the rich new physics these textures possess [1-4]. These spin textures were first observed in bulk magnets lacking inversion symmetry at low temperatures [5-8]. In these materials, the broken symmetry gives rise to a skew exchange interaction, commonly known as the Dzyaloshinskii–Moriya interaction (DMI) [9, 10], that stabilizes a helical spin structure at remanence. Under the application of a perpendicular magnetic field, the helical spin configuration undergoes a magnetic phase transformation into a hexagonal skyrmion lattice [5-8]. Their topological nature enables unique properties such as current-driven motion with low current density [11, 12], insensitivity to defects [13, 14], topological Hall effect [15-17], skyrmion Hall effect [18-20], and non-trivial skyrmion spin wave dynamics resulting from microwave perturbations [21-23]; among other characteristics. As a result, these textures have been suggested as a building block for next generation non-volatile memory and logic devices [13, 24-27].

There are increasing research efforts to develop thin-film hetero-structured materials that form skyrmions at room temperature and are controllable with electrical currents. The observation of a sizeable interfacial DMI resulting from interfacing thin ferromagnetic films with a non-magnetic layer that possesses high spin-orbit coupling (*e.g*. 5-*d* transition metals such as Ta, W, Pt, Ir, *etc*.) [28-33] provides a pathway to engineer material properties to form skyrmions in thin films. Trilayer ferromagnetic heterostructures exploiting interfacial DMI and asymmetric interfaces are particularly attractive materials systems. Examples of reported materials systems where skyrmions are observed include Ta/CoFeB/TaOx [33], Ta/CoFeB/MgO [34], Pt/Co/Ta [34], and Pt/Co/Ir multilayers [35]. This approach has produced most of the materials showing room-temperature skyrmions with the exception of a few crystal plates [36-39] and topologically-equivalent dipole-stabilized skyrmions [40-42]. So far, the tuning of interfacial DMI has mostly focused on 5-*d* transition metals do to their high atomic number Z and corresponding high spin-orbit coupling.

Here, we report the observation of DMI-stabilized skyrmions in Pt/Co/Os/Pt thin-film heterostructures. Os is a 5-*d* transition metal located between Re and Ir but there are no reports on contributions to interfacial anisotropy, DMI or skyrmion formation in thin-film heterostructures. We find that adding a thin Os layer (~ 0.2 nm) into a Pt/Co/Pt structure between the Co and top Pt layer dramatically alters the magnetic properties enabling the formation of room-temperature skyrmions. We observe the skyrmions exist in narrow materials composition, magnetic field and temperature window and are highly thermally active. The skyrmion features (average diameter ~2.3 µm) are of comparable size to those reported for heavy-metal broken symmetry trilayer thin-film structures [33,43]. The skyrmions can be moved with low current densities (~$4 \times 10^4$ A/cm$^2$) and we observe the skyrmion Hall effect when moving the skyrmions with relatively low current density (~$10^6$ A/cm$^2$) comparable to other DMI skyrmions that form in ferromagnetic trilayers [33,43].

The ferromagnetic Pt/Co/Os/Pt thin-films are sputter deposited at ambient temperature in a 3-mTorr Ar pressure. The sputter system base pressure was < $3 \times 10^{-8}$ Torr. The films investigated have the following structure: SiOx/Ta(5 nm)/Pt(3 nm)/Co(1.2 nm)/Os($t_{Os}$)/Pt(3 nm)/Ta(5 nm) in which the Os layer thickness is varied from $t_{Os} = 0$ nm to 0.4 nm. The samples were deposited on Si substrates with a native oxide layer for magnetic characterization and Si/SiO$_x$ substrates with a 300-nm thermal oxide layer for transport measurements. Magnetic hysteresis measurements were characterized using a Quantum Design Versalab equipped with a vibrating sample magnetometer and magneto-optical Kerr effect (MOKE) hysteresis measurements. The field- and temperature-dependent domain morphology was imaged with by MOKE microscopy using an Evico Magnetics microscope utilizing a Zeiss 20X and 50X magnifying lenses with a spatial resolution of 660 and 250 nm/pixel, respectively.

Figure 1 highlights the room-temperature magnetic properties and domain morphology at zero-field of Pt/Co/Os($t_{Os}$)/Pt heterostructures for $t_{Os}$ = 0, 0.1, 0.2 and 0.4 nm. The room-temperature hysteresis loops are shown in Figs. 1a-d. For the reference Pt(3 nm)/Co(1.2 nm)/Pt(3 nm) structure (Os thickness $t_{Os} = 0$ nm) we observe the expected square loop with full remanence and a coercivity field of $H_c = 164$ Oe (Fig. 1a). This is consistent with numerous previous studies of Pt/Co/Pt structures [44]. For Pt/Co/Os($t_{Os}= 0.1$ nm)/Pt, the magnetic hysteresis loop continues exhibiting a square loop with full remanence (Fig. 1b) but the coercive field decreases to $H_c = 150$ Oe with a slight decrease in the saturation magnetization ($M_S = 950$ emu/cm$^3$). Increasing the Os thickness to $t_{Os} = 0.2$ nm (Fig. 1c) which corresponds to roughly one monolayer results in a dramatic reduction of the room-temperature coercive field to $H_c \sim 1$ Oe and saturation magnetization ($M_S = 290$ emu/cm$^3$). Further increasing the Os thickness to $t_{Os} = 0.4$ nm produces a spin-reorientation transition of the magnetization from perpendicular to easy-plane anisotropy and a further reduction of the saturation magnetization. The out-of-plane magnetic loop appears as a hard-axis and the sample saturates in a field of $H_k = 500$ Oe (Fig. 1d).

To further understand the effects of introducing an Os interface, we inspected the perpendicular domain morphology of Pt/Co/Os($t_{Os}$ = 0, 0.1, 0.2, 0.4 nm)/Pt heterostructures using polar-MOKE microscopy at room temperature (Fig. 1e-h). The magnetic domains are represented by variations of intensity that result from small alterations of the polarization when the incident light is reflected off the magnetic domains. Figures 1e and 1f shows a snapshot of a magnetic domain at zero-field for Os thickness $t_{Os}$ = 0 nm and $t_{Os}$ = 0.1 nm that was captured by slowly approaching the coercivity field and then reducing the field to remanence. In such a procedure, a single domain is trapped in the image showing that the field reversal is controlled by domain wall motion and the characteristic domain size is large as expected for thin Pt/Co/Pt films [44]. For Pt/Co/Os($t_{Os}$ = 0.2 nm)/Pt we observe a dramatic decrease in the domain size where the remanent magnetization state consists of labyrinth stripe domains with an average domain periodicity of 4.4 µm. These small domain features are usually observed in [Pt/Co]$_{xN}$ multilayers with larger $N$ repetitions [44, 45] and are not characteristic of thin Pt/Co/Pt films. Typically, magnetic domain morphologies are formed in thicker films because there is higher magnetostatic energy $2\pi M_S^2$ and in order to minimize the total magnetic energy, magnetic domain walls are introduced which results in perpendicular magnetic domains of varied shape and size [46, 47]. Here, the observation of these small domains suggests there is an additional energetic contribution that lowers the domain wall energy. As will be discussed, this suggest the presence of DMI is at least partly responsible for the formation of these small domain features. Increasing the Os thickness to $t_{Os}$ = 0.4 nm results in an image with uniform illumination across the field of view consistent with the magnetization being in the easy plane with no perpendicular magnetic domains present (Fig. 1h).

Figure 2 shows the room-temperature field-dependent domain morphology of the Pt/Co/Os ($t_{Os}$ = 0.2 nm)/Pt sample. These images are captured as a perpendicular magnetic field is applied from zero-field towards magnetic saturation. At zero-field, the domain morphology consists of disordered stripe domains (Fig. 2a). The enclosed region in Fig. 2a is enlarged in Fig. 2b. We also observe the disordered stripe domains fluctuate in real-time (Supp. Movie 1), a characteristic state that has been observed when the magnetic system is close to a magnetization spin-reorientation transition [48] or the Curie temperature [49, 50]. As the magnetic field is increased, the extremities of the stripe domains begin to collapse into cylindrical-like magnetic features. We identify these cylindrical magnetic domains as DMI skyrmions based on their response to an applied current. At $H_z$ = 0.7 Oe the domain morphology consists of a closed-packed mixture of disordered stripe domains and skyrmions (Fig. 2c, d). Increasing the field, the remaining stripe domains will collapse to form a closed packed lattice of skyrmions (Fig. 2e, f). The skyrmion lattice does not have long range order as commonly observed in bulk crystal magnets where Bloch-type DMI skyrmions have been reported [5-8]. We also observe the diameter of the skyrmion features does not vary significantly over the field of view (Fig. 2e, h). Typically, ferromagnetic heterostructures with heavy metal broken symmetry result in DMI skyrmions with varied diameter [34,35,43,51] and this has been correlated to non-uniformity of the heavy metal interface [51,52]. In our case, we can ascertain the effects of the Os interface is relatively uniform across the surface of the film. In addition, we also observe real-time thermal fluctuations of the skyrmion domains (Supp. Movie 2). As the magnetic field is further increased, the skyrmions begin to collapse as we approach magnetic saturation. At $H_z$=1.8 Oe, the domain morphology consists of a disordered skyrmion phase (Fig. 2g, h). Similar to other interfacial DMI skyrmion materials [33,43], the Pt/Co/Os ($t_{Os}$ = 0.2 nm)/Pt film require low magnetic fields ($H_{sat}$ < 5 Oe) to saturate.

Figures 3 and 4 highlight the temperature dependence of the magnetic properties. The temperature dependence of $M_S$ for Pt/Co/Os($t_{Os}$ = 0, 0.1, 0.2, 0.4 nm)/Pt heterostructures are shown in Fig. 3. As seen in Fig. 1 the introduction of Os interface results in dramatic reduction of $M_S$ that persists for all temperatures and there is a corresponding reduction of the Curie temperature. The symmetric Pt/Co/Pt and the Pt/Co/Os($t_{Os}$ = 0.1 nm)/Pt heterostructure exhibits a magnetization that does not vary greatly with temperature between the 50K and 375K suggesting a Curie temperature well above room temperature. As the Os thickness is further increased to $t_{Os}$ = 0.2 nm, $M_S$ drops by over half compared to the symmetric Pt/Co/Pt film and a somewhat lower $M_S$ is observed for Os thickness of $t_{Os}$ = 0.4nm (Fig. 1e). Also there is a dramatic reduction the Curie temperature to below 400 K for $t_{Os}$ ≥ 0.2 nm.

Figure 4 shows the temperature and magnetic-field-dependent measurements of the Pt/Co/Os($t_{Os}$ = 2nm)/Pt sample near room temperature where the skyrmion phase is observed. For the images the magnetic field is applied after positive magnetic saturation. At T = 295 K, the perpendicular magnetic hysteresis exhibits a loop with full remanence and a coercive field $H_c$ = 0.65 Oe. As expected for a nearly square loop the domain reversal revolves via the nucleation and domain growth as seen in Fig. 4b. Further decreasing the field, $H_z$ = -1.3 Oe, we observe the domain morphology consists of disordered stripe domains at the tail of the magnetic loop (Fig. 4c). At T = 297 K, the hysteresis loop no longer shows full remanence and magnetic reversal begins at a positive nucleation field of $H_n$ = 0.25 Oe. Near the coercive field, $H_z$ = -0.4 Oe, the film forms perpendicular stripe domains as shown in Fig. 4d. Decreasing the field to $H_z$ = -0.8 Oe, causes the extremities of the stripe domains to contract and their width to decrease (Fig. 4f). As the magnetic field is further increased the stripes will eventually collapse into disordered cylindrical magnetic domains before the film saturates. The latter is the typical field-dependent evolution of perpendicular magnetic stripes [46,53,54]. At T = 299 K, the magnetic loop continues to exhibit characteristics associated with the presence of perpendicular stripe domains, but now the nucleation field $H_n$ has shifted to a higher positive field, $H_n$ = 1 Oe (Fig. 4g). The field dependent domain morphology is comparable to the one previously described in Fig. 2: at remanence the film exhibits disordered stripe domains, while under positive and negative perpendicular fields, three different magnetic states are observed: coexisting stripes and skyrmions (Fig. 4h), skyrmion lattice (Fig. 4i), and then disordered skyrmions as highlighted in Figs. 2g and h. At T = 301 K, the magnetic loop has features associated with magnetization along the easy plane, yet the hysteresis is not closed and the film continue to exhibit a nucleation field at $H_n$ = 2 Oe (Fig. 4j). The field-dependent domain morphology is comparable to 299K, but consists of smaller magnetic features, such as disordered stripe domains with a periodicity of 3-μm at zero field and 2.5-μm skyrmion features at $H_z$ = -1.4 Oe. We also find the closed-packed lattice exists for a wider range of magnetic fields, $H_z$ = -0.9 Oe to $H_z$ = -1.6 Oe. At T = 303 K, the magnetic loop suggests the magnetization now lies in the plane of the film given the nucleation field has become the saturation field and the loop does not exhibit hysteresis (Fig. 4m). Conversely, there is no evidence of perpendicular magnetic domains as a function of applied field (Fig. 4n, o).

    As seen in Fig. 4 the emergence of the skyrmion phase is extremely sensitive to temperature and magnetic fields. Because of this sensitivity we also observe that small variations in the sample structure can shift the transition. Samples with nominally the same structure can have the skyrmion phase region shift in temperature. However, for each sample we observe the skyrmion phase near the spin-reorientation transition from out-of-plane to in-plane magnetization with temperature. The appearance of small domains near the spin reorientation has been observed in a number of magnetic thin film systems either from changing temperature or changing film thickness [55-58]. At the transition where the shape and crystalline anisotropy are balanced, the average domain size tends to decrease dramatically and in some cases small bubble domains can be observed [57]. Thus the presence of the bubbles observed in the microscope image does not necessarily mean they are

skyrmions or have a preferred chirality. In magnetic films or crystal platelets, the topology of magnetic skyrmions can be verified by directly imaging the spin textures with Lorentz transmission electron microscopy, a real-space imaging technique [7,58]. Alternatively, one can study the response of the spin textures to currents to determine if they possess a uniform chiral topology. Since skyrmions possess a topological charge, these textures will displace along a transverse direction relative to the current driving direction resulting from the topological Magnus force. The transverse displacement of the skyrmions is commonly known as the skyrmion Hall effect [18-20].

To demonstrate the cylindrical-like magnetic features observed in Pt/Co/Os/Pt films are, in fact, DMI skyrmions with a common chirality, we patterned continuous films into devices with 20-μm wide wires (Supp. Fig. 1, insert). The cylindrical-like magnetic domains are stabilized under the application of a perpendicular magnetic field, $H_z = 0.9$ Oe, and a d.c. electric current is applied along the length of the wire. We observe the skyrmion move the direction of the current flow and start to move at current densities as low as $2 \times 10^4$ A/cm$^2$ and the average speed initially increases linearly with current (Supp. Fig. 1). The fact that the skyrmions all move in the direction of the current flow suggests a common chirality. However, we were not able to observe the skyrmion Hall effect within the wire because of the temporal resolution of the microscope and the limited wire width. To observe the manifestation of the skyrmion Hall effect we monitored the skyrmion motion from a wire into an extended film. Figure 5 shows polar-MOKE microscopy snapshots of the skyrmion domains at different time intervals *t* while a constant current density of amplitude $J = 1 \times 10^6$ A/cm$^2$ within the wire flows along the patterned wire into the full film. After saturating in a positive field and applying a small negative out-of-plane field, the magnetic domains appear as white-contrast features and within the field-of-view there exists a combination of cylindrical-like magnetic domains and stripe domains at $t = 0$ s (Fig. 4a). When the electric current flows through the wire, we observe the magnetic features begin to flow in the direction of the current flow and then emerge into the full film (Supp. Movie 3). Since the current density is much higher in the wire, the domain features are displaced at higher speeds within the wire than in the full film and we do not have the temporal resolution to track the skyrmions in the wire. However, it is observed that the cylindrical features exiting the wire move transverse to the direction of the current flow and these textures begin to accumulate on the left-hand side (LHS) of the full film. Snapshots of the time-varying domain morphology at $t = 5$ s (Fig. 4b), $t = 12$ s (Fig.

4c) and $t = 21$ s (Fig. 4d) show the population of skyrmions textures accumulated on the LHS increases and no cylindrical textures populate the right-hand side (RHS) of the full film. After saturation in a negative field and then applying a small positive magnetic field, the cylindrical magnetic domains appear as dark-contrast features (Figs. 4e-h) in the image. These features also displace transverse to the current direction; however, these textures accumulate on the RHS of the full film and the population of these textures increases with time (Figs. 4f-h). The difference in transverse motion for opposite magnetization spin textures suggests these domains are, in fact, DMI stabilized skyrmions with a fixed chirality [19, 20]. Since the Magnus force is given by $\vec{G} \times \vec{v}$, where the gyromagnetic coupling vector $\vec{G} = (0, 0, -4\pi Q)$ and $\vec{v}$ is the skyrmion velocity, we can infer that the skyrmions topological charge from the transverse motion. Based on the skyrmion displacement relative to the current flow and the polarity of the perpendicular magnetic field infer that the skyrmions that form under negative fields possess a topological charge of $Q = -1$ (Figs 4a-d); conversely, positive fields results in skyrmions with $Q = +1$ (Figs. 4e-h) [19, 20, 43]. From this we can determine that the sign of the DMI is negative if arising from the bottom Pt/Co interface which is in agreement with previous measurements [28, 60]. Since the motion of skyrmions is in the direction of the current flow we cannot ascertain whether the mechanism driving the displacement is spin-orbit-torque but we believe the motion results from the SOT arising from the bottom Pt layer in agreement with previous studies of Pt/Co-based heterostructures [61, 62].

In conclusion we have explored the magnetization, anisotropy and formation of DMI skyrmions in Pt/Co/Os/Pt heterostructures that form in narrow composition, temperature and magnetic field window. The skyrmion features appear in an energetic region near the temperature driven magnetization spin reorientation from out-of-plane to in-plane anisotropy. We have also shown that these skyrmion features can be displaced with low current densities that are comparable to those used to displace DMI skyrmions that form in non-centrosymmetric bulk magnets and display a skyrmion Hall effect.

The research was supported by the University of California Multicampus Research Programs and Initiatives (MRPI) award number MRP-17-454963 on Electrical Control of Topological Magnetic Order.

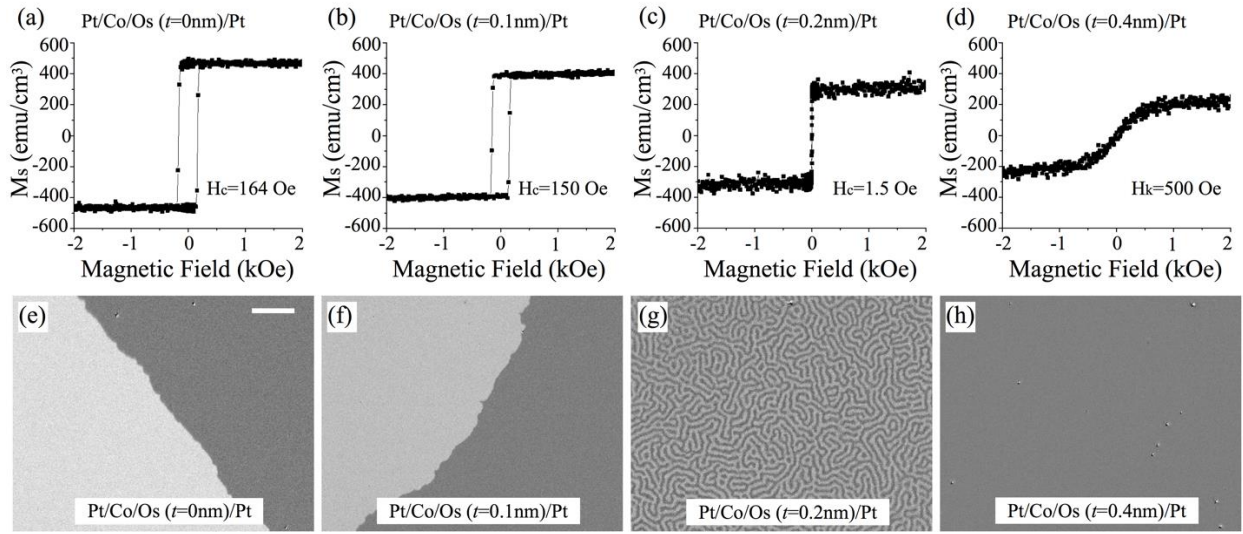

**Figure 1:** (a-d) The magnetic hysteresis loops for Pt/Co/Os($t_{Os}$)/Pt films with (a) $t_{Os}$ = 0 nm, (b) $t_{Os}$ = 0.1 nm, (c) $t_{Os}$ = 0.2 nm, and (d) $t_{Os}$ = 0.4 nm measured at T = 300K under an out-of-plane magnetic field. (e-h) The domain morphology at remanence for the Pt/Co/Os($t$)/Pt films with (e) $t_{Os}$ = 0 nm, (f) $t_{Os}$ = 0.1 nm, (g) $t_{Os}$ = 0.2 nm, and (h) $t_{Os}$ = 0.4 nm is obtained with polar-MOKE microscopy at T = 300K. The scale bar in (e) corresponds to 25μm.

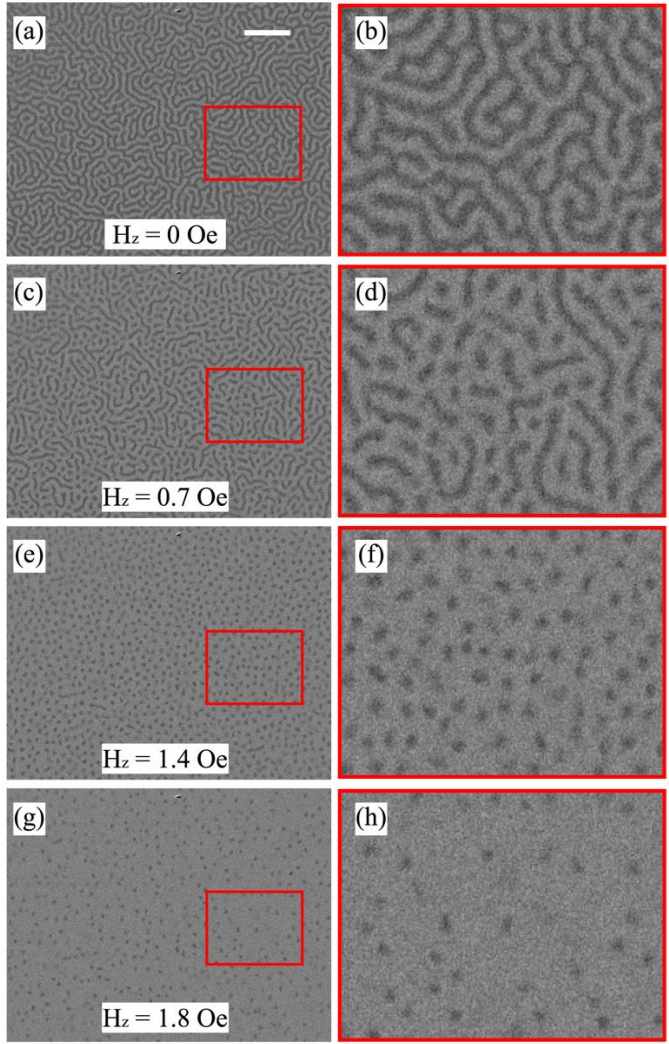

**Figure 2**: Field dependent domain morphology of Pt/Co/Os($t_{Os}$ = 0.2 nm)/Pt at 300K. The images are captured as a perpendicular magnetic field is applied from zero-field towards positive magnetic saturation. The black features depict magnetization ($-m_z$) and the gray regions represent magnetization ($+m_z$). Different magnetization states are observed as a function of applied magnetic field: (a) disordered stripe domains, (c) coexisting stripe and skyrmions, (e) closed packed skyrmion lattice and (g) disordered skyrmions. The enclosed regions in (a, c, e, g) are enlarged to detail domain morphology in (b, d, f, h). The scale bar in (a) corresponds to 25µm.

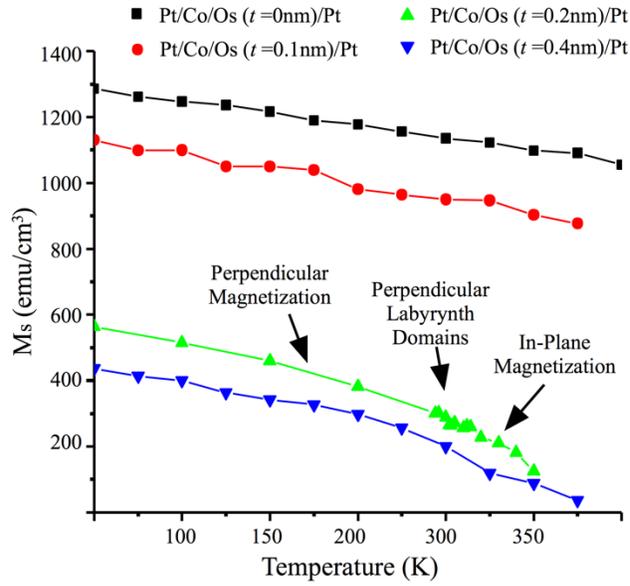

**Figure 3:** The temperature dependence of the magnetization for the Pt/Co/Os($t_{Os}$ = 0, 0.1, 0.2, 0.4nm)/Pt films is detailed from 375K to 50K. Based on polar-MOKE microscopy measurements on Pt/Co/Os($t_{Os}$ = 0.2 nm)/Pt, we have identified temperature regions that favor specific remanent domain states.

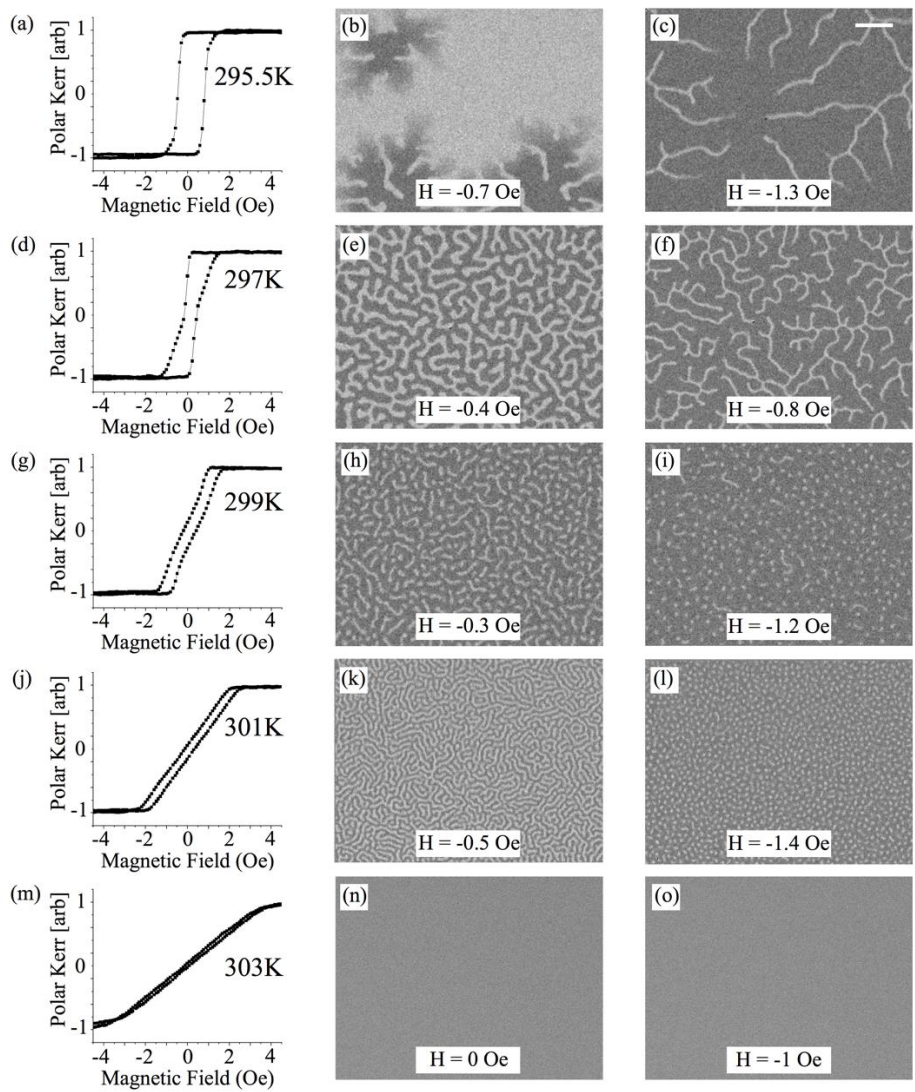

**Figure 4**: Temperature dependence of the polar-Kerr magnetic hysteresis and field-dependent domain morphology of Pt/Co/Os($t_{Os}$ = 0.2 nm)/Pt at (a-c) T = 295 K, (d-f) T = 297 K, (g-i) T = 299K, (j-l) T=301 K and (m-o) T=303 K. Images of the magnetic domain states are captured as the magnetic field is applied from positive to negative magnetic saturation. The scale bar in (c) corresponds to 25 μm.

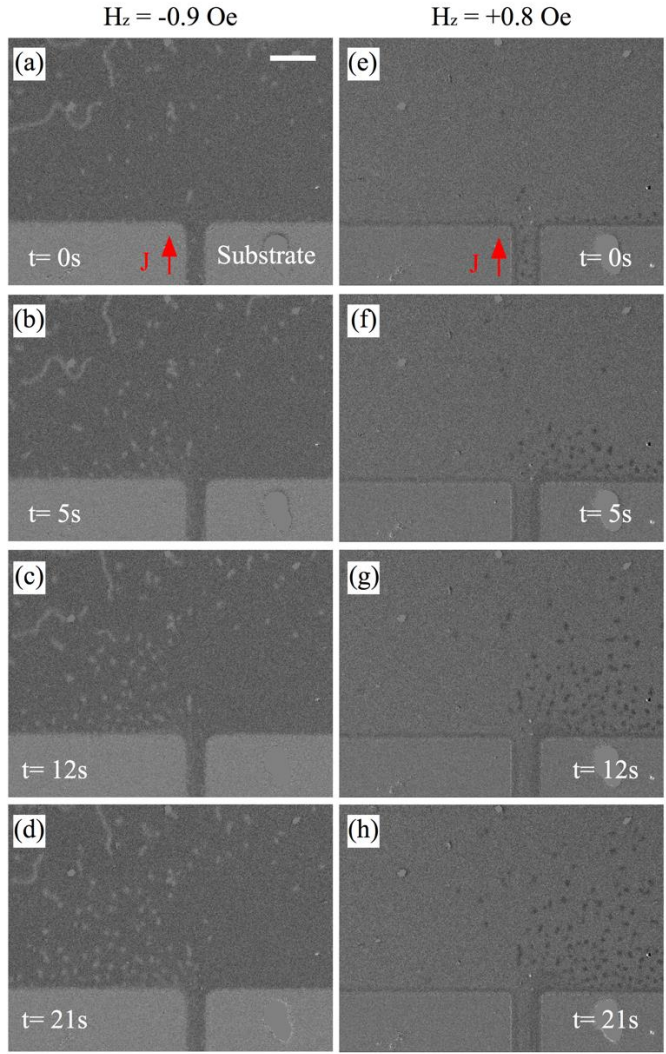

**Figure 5:** Domain morphology snapshots of the current-induced displacement of skyrmion textures in Pt/Co/Os($t_{Os}$ = 0.2 nm)/Pt at different time intervals (t = 0, 5, 12, and 21 s) after the current is turned on. The dynamics are detailed for both perpendicular magnetization polarizations: (a-d) white contrast textures detail the (+$m_z$) magnetization domains, whereas (e-h) black contrast show the (-$m_z$) magnetization domains. The skyrmion textures are pushed along a 20μm wire, shown at the bottom portion each snapshot, with fixed current density $J = 1 \times 10^6$ A/cm² into the full film. The scale bar in (a) corresponds to 25μm.